\documentclass{article}
\usepackage{spconf,graphicx}
\usepackage{amsfonts, amsmath, amssymb, amsthm, bbm, bm}
\usepackage{algorithm}
\usepackage{algpseudocode}
\usepackage{nicematrix}
\usepackage{diagbox}
\floatname{algorithm}{Method}
\usepackage{enumitem}
\setlist{nosep, leftmargin=14pt}

\usepackage{mwe} 
\usepackage{array}
\usepackage{multirow}
\usepackage{nicematrix}
\usepackage[breaklinks=true,colorlinks=true,bookmarks=false]{hyperref}

\DeclareMathOperator*{\argmin}{argmin}

\title{Optimal Transport Guided Unsupervised Learning for Enhancing low-quality Retinal Images}
\name{Wenhui Zhu$^1$, Peijie Qiu$^2$, Mohammad Farazi$^{1}$, Keshav Nandakumar$^{3}$, Oana M. Dumitrascu$^4$, Yalin Wang$^1$}
\address{
$^1$ School of Computing and Augmented Intelligence, Arizona State University, AZ 85281, USA \\
$^2$ McKeley School of Engineering, Washington University in St. Louis, St. Louis, MO 63130, USA \\
$^3$ Honors’ Barret College, Arizona State University, AZ 85281, USA \\
$^4$ Department of Neurology, Mayo Clinic, Scottsdale, AZ 85251, USA}
\begin{document}
%
\maketitle
\begin{abstract}
Real-world non-mydriatic retinal fundus photography is prone to artifacts, imperfections and low-quality when certain ocular or systemic co-morbidities exist. Artifacts may result in 
 inaccuracy or ambiguity in clinical diagnoses. In this paper, we proposed a simple but effective end-to-end framework for enhancing poor-quality retinal fundus images. Leveraging the optimal transport theory, we proposed an unpaired image-to-image translation scheme for transporting low-quality images to their high-quality counterparts. We theoretically proved that a Generative Adversarial Networks (GAN) model with a generator and discriminator is sufficient for this task. Furthermore, to mitigate the inconsistency of information between the low-quality images and their enhancements, an information consistency mechanism was proposed to maximally maintain structural consistency (optical discs, blood vessels, lesions) between the source and enhanced domains. Extensive experiments were conducted on the EyeQ dataset to demonstrate the superiority of our proposed method perceptually and quantitatively. 
\end{abstract}
\section{Introduction}
Not only retinal fundus images can assist ophthalmologists to diagnose ocular diseases, but they also serve automated computer-aided systems for diagnosing and monitoring diabetes and even Alzheimer’s disease~\cite{cheung2022deep}. Both human and computer-aided diagnoses prefer operating on high-quality retinal images.
The retinal fundus imaging systems, especially nonmydriatic ones, however, inherently face shading artifacts and blurring because of light transmission disturbance, defocusing, or suboptimal human operations \cite{shen2020modeling}, resulting in degradation of imaging quality. 
Such degradation, e.g., occlusion of blood vessels, lesions, leads to failure or inaccuracy in the final diagnosis.
Enhancing low-quality retinal images into high-quality counterparts is, therefore, inevitable for many downstream tasks, directly or indirectly related to diagnosing various retinopathies or automated analyses. 

The process of mapping low-quality retinal images to their high-quality counterparts, where we defined the domain consisting of enhanced images as enhanced domain, is usually modeled as an end-to-end image-to-image translation task. Many previous explorations~\cite{https://doi.org/10.48550/arxiv.1606.07536, pix2pix, cyclegan, https://doi.org/10.48550/arxiv.1612.05424} in this task leverage generative adversarial networks(GANs) where the adversarial training progressively leads us to the photo-realistic renderings. The key idea is to map a source domain $\mathcal{Y}$ to a target domain $\mathcal{X}$, which maps the source distribution to a target distribution $\mathbb{P}_{\mathcal{Y}} \rightarrow \mathbb{P}_{\mathcal{X}}$. This mapping becomes more challenging when the input and target are unpaired because no direct groundtruth data are available. To reduce the searching space of such mappings, \cite{https://doi.org/10.48550/arxiv.1606.07536, https://doi.org/10.48550/arxiv.1612.05424} provide a task-specific regularization, while the CycleGAN~\cite{cyclegan} generalizes them as a cycle consistency. However, as a general image generation model, CycleGAN has several drawbacks when applied to retinal image generation. First, it is burdened with its expensive computation with concurrent training of a pair of generators and discriminators. Secondly, it may lose lesions and introduces non-existing vessels. For example, Wang et al.~\cite{9763342} proposed an optimal transport GAN for unsupervised natural image denoising. From our experiments, we found a direct application of their method led to the destruction of vessel and lesion structures. Different from natural images with additive noises, the degradation of retinal images is more complicated and,  therefore, more challenging to model. 

 Inspired by \cite{9763342}, we proposed an alternative view of image-to-image translation with a standpoint of optimal transport theory to enforce the consistency between the enhanced domain and our target domain (high-quality images) while preserving the information between the source and target domains to prevent lesion tampering and the generation of unrealistic non-existing blood vessels.
The contributions of our work can be summarized in three aspects: (1) We proposed an optimal transport-guided \emph{domain consistency} that ensures the consistency between the enhanced domain and our target domain. (2) A unified GAN-based unsupervised retinal image enhancement training scheme was introduced.  (3) To mitigate the inconsistency of vessels, optic disc, and lesions between before and after enhancements, we introduced a maximally information-preserving consistency loss in conjunction with a data resampling mechanism. 


\section{Methods}
\begin{figure*}[!t]
\centering
\includegraphics[width=\textwidth]{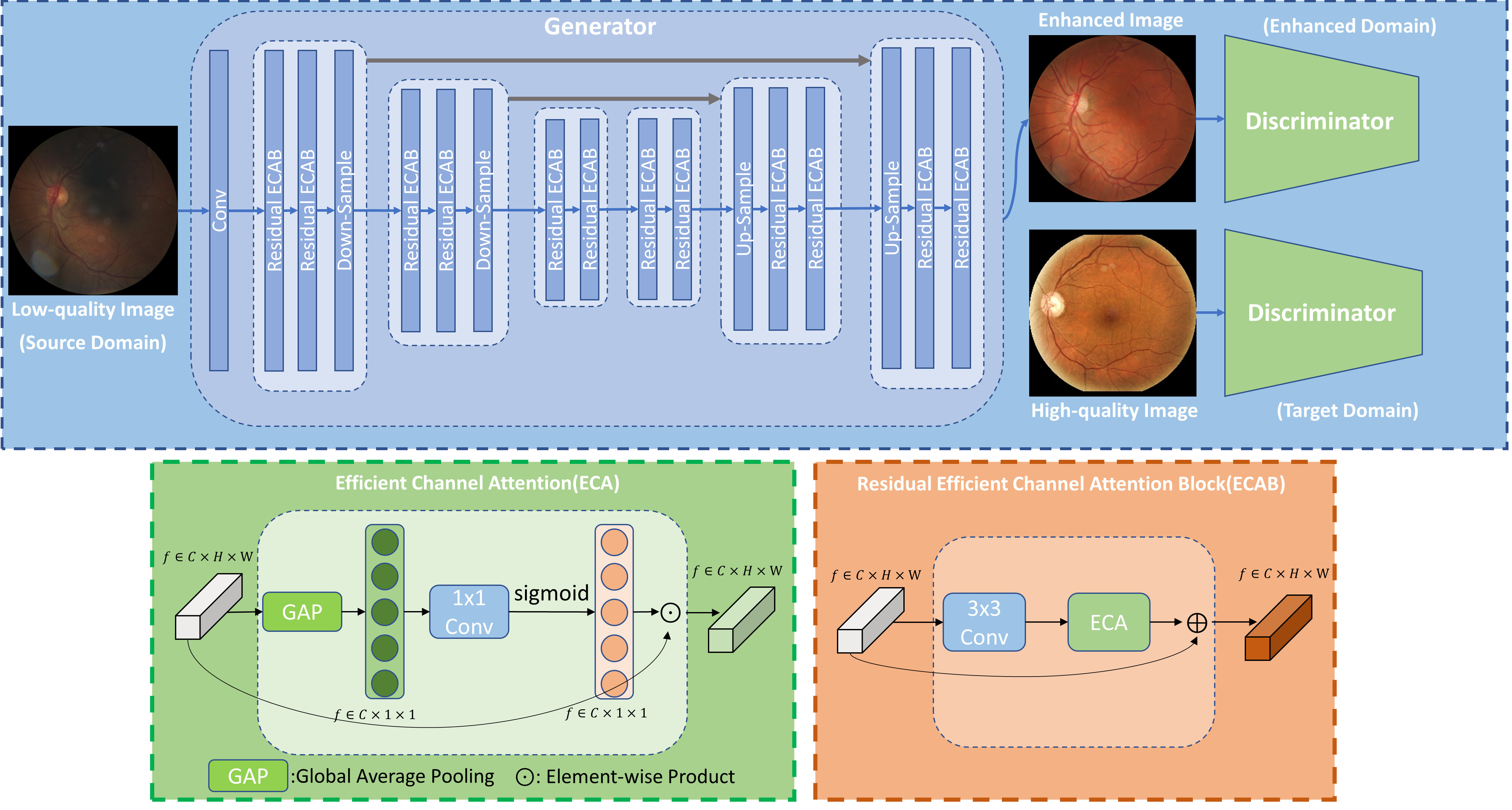}
\caption{Network Architecture of the proposed method. The Efficient Channel Attention is added for each residual block in the generator to enhance the lesion representation, and the discriminator is adapted from the one in \cite{disc}.}
\label{fig:network}
\end{figure*}

Our proposed method consists of three modules: optimal transport-guided domain consistency, maximally information-preserving consistency, and refined data resampling for lesion consistency. The entire framework of our proposed method is shown in Fig.\ref{fig:network}.



\subsection{Optimal Transport Guided Domain Consistency}
Let $  \mu \sim \mathbb{P}_{\mathcal{X}}$  and $\nu \sim \mathbb{P}_{\mathcal{Y}}$ denote two probability measures on the input and target probability spaces, respectively. The \emph{Monge’s} optimal transport map $G: \mathcal{Y} \rightarrow \mathcal{X}$ that minimizes the cost of transporting $\nu$ to  $\mu$ is given by
\begin{equation}\label{eqn:Monge}
    \begin{split}
        G^* = \inf \limits_{\substack{ f: \mu(X) =  \nu(G^{-1}(X))  } } \int_{\mathcal{Y}} \mathcal{L}_c(Y, G(Y))  \; \mathrm{d} \nu(Y) \ \\
    \end{split}
\end{equation}
where $\mathcal{L}_c: \mathcal{Y} \times \mathcal{X} \rightarrow [0, +\infty]$ denotes the cost function. By parameterizing the optimal transport map as a neural network $G_{\theta}$, Equation~\ref{eqn:Monge} can be discretized to
\begin{equation}\label{eqn:discreteOT}
    \begin{split}
        \theta^* = \argmin \limits_{\theta: \ \mathbb{P}_{\mathcal{X}} =  \mathbb{P}_{G_{\theta}(Y)} } \mathbb{E}_{Y \sim \mathbb{P}_{\mathcal{Y}}} [ \mathcal{L}_c(Y, G_{\theta}(Y)) ].
    \end{split}
\end{equation}
It is worth noticing that Equation~\ref{eqn:discreteOT} yields an unsupervised training scheme where the minimization is decoupled with the high-quality images $X$, and the constraint can be achieved by unpaired adversarial training. The constrained optimization can be further relaxed to an unconstrained optimization by applying \emph{Lagrange} multiplier to Equation~\ref{eqn:discreteOT} yielding
\begin{equation}\label{eqn:LagrangeOT}
      \theta^* = \argmin \limits_{\theta} \mathbb{E}_{Y \sim \mathbb{P}_{\mathcal{Y}}} [ \mathcal{L}_c(Y, G_{\theta}(Y)) ] + \lambda d(\mathbb{P}_{\mathcal{X}}, \mathbb{P}_{G_{\theta}(Y)})
\end{equation}
where $d$ measures the divergence between two distributions with $d(:, :) \leq 0$. To take advantage of GANs, we achieve the divergence constraint $d(:, :)\leq 0 $ adversarially by optimizing the Wasserstein-1 distance~\cite{NIPS2017_892c3b1c} given by
\begin{equation}
\begin{split}
    \mathcal{W}_1(\mathbb{P}_{\mathcal{X}}, \mathbb{P}_{G_{\theta}(Y)}) = \sup \limits_{||D_{w}||_L \leq 1} \mathbb{E}_{X \sim \mathbb{P}_{\mathcal{X}}} [D_{w}(X)] - \\
    \mathbb{E}_{Y \sim \mathbb{P}_{\mathcal{Y}}} [D_{w}(G_{\theta}(Y))]
\end{split}
\end{equation}
where $D_{w}$ denotes the discriminator parameterized by $w$ with a 1-Lipschitz constraint which is approximated by Gradient Penalty~\cite{NIPS2017_892c3b1c} in our experiments.

Equation~\ref{eqn:LagrangeOT} implies the consistency between the source and a target domain is traded off by the consistency between the target domain and the enhanced domain, which matches our initial expectation that the distribution of the enhanced low-quality images should align with that of the high-quality images while having the same underlying structures as the low-quality images. 
Low-quality enhancement could be fully achieved by training a GAN with generator $G_{\theta}$ and discriminator $D_{w}$ via optimizing the following objective function:
\begin{equation}
     \max \limits_{G_{\theta}} \min \limits_{D_w} \ \mathbb{E}_{Y \sim \mathbb{P}_{\mathcal{Y}}} [ \mathcal{L}_c(Y, G_{\theta}(Y)) ] + \lambda \mathcal{W}_1(\mathbb{P}_{\mathcal{X}}, \mathbb{P}_{G_{\theta}(Y)})
\end{equation}
       

\subsection{Maximal Information Consistency}
Next, we will introduce how to choose the loss function $\mathcal{L}_c$ to enforce data consistency for our task. 
When $\mathcal{L}_c$ is convex, the strong duality between Equation~\ref{eqn:LagrangeOT} and Equation~\ref{eqn:discreteOT} holds, meaning they achieve identical optimality.
Conventionally, the $L_1$ or $L_2$ norm is a common convex choice to enforce the data consistency, where $L_1$ norm leads to the optimal median while $L_2$ norm leads to the optimal mean \cite{n2n} from a statistical point of view. But either $L_1$ or $L_2$ norm will result in blurring in rendered images because of the smoothness of sharp edges and loss of high-frequency local structures \cite{pix2pix}, which is undesired in our task. From our early experiments with CycleGAN~\cite{cyclegan} using $L_1$ norm as a consistency loss, we observed some pathologically meaningful structures, particularly lesions in diabetic retinopathy, were lost in the enhancement, as shown in Fig.~\ref{fig:non_ref}. To preserve more local structures, particularly those that are pathologically meaningful, we choose Multi-Scale Structural Similarity Index Measure $\mathbf{SSIM}_{MS}$ \cite{wang2004image, wang2003multiscale} as our consistency loss, which is given by
\begin{equation}
    \mathcal{L}_c(Y, G_{\theta}(Y)) = 1- \mathbf{SSIM}_{MS}(Y, G_{\theta}(Y)),
\end{equation}
where $G_{\theta}(Y)$ is the rendered high-quality images. The SSIM consistency loss is locally quasi-convex\cite{6059504} minimizing the duality gap between Equations~\ref{eqn:LagrangeOT} and \ref{eqn:discreteOT} while maximizing the mutual information between the source domain and enhanced domain. 

In our task, the enhanced high-quality images from their low-quality counterparts, in general, have the same underlying structures, e.g., blood vessels, lesions, optical disks, etc. With this observation, we design a specific U-shape ~\cite{ronneberger2015unet} generator, enabling the low-level semantic information flow from low-quality images to their high-quality enhancements. To further augment the performance of lesion preservation, we add an efficient channel attention block to each residual block in the generator. Because different types of diabetic lesions have different sensitivity in different image channels. 

\subsection{Lesion Consistency via Refined Resampling}
Despite our maximal mutual information consistency loss, facilitating lesion consistency between the low-quality images and their enhancements is still a challenging task due to the variation in the occurrence of lesions. In diabetic retinopathy, images at different diabetic retinopathy levels consist of different types of lesions. Even images at the same diabetic retinopathy level are likely to consist of different lesions, e.g., in proliferative diabetic retinopathy, the lesion may include hard exudates, hemorrhages, and microaneurysms. To mitigate this issue, we take advantage of prior knowledge of the lesions for sampling input-target pairs by ensuring the input and target are at the same diabetic retinopathy level. Not only does doing so guide the lesion consistency between the low-quality images and their enhancements, but it also ensures that the distribution of lesions in the enhanced images is close to that in the low-quality images.  




\section{Experiments}
We evaluated our method through two major experiments: No-Reference Quality Evaluation and Full-Reference Quality on the EyeQ dataset \cite{10.1007/978-3-030-32239-7_6}. The main comparisons are with the popular unsupervised image-to-image translation and noise reduction adversarial generation models, including (1) CycleGAN \cite{cyclegan}, and (2) optimal transport-based method (OTT-GAN)~\cite{9763342}.

\subsection{Dataset}

The EyeQ \cite{10.1007/978-3-030-32239-7_6} dataset consists of 9239 training images and 11362 testing images. The original dataset is manually labeled into three quality levels: good, usable, and reject. Our goal is to convert all reject images to high-quality images (good). All models were trained on the official training set: 6342 good images and 1544 reject images and evaluated on the official testing dataset: 5966 good and 2195 reject images. 

\subsection{Implementation Details}
Our proposed method was implemented in \emph{PyTorch}. For training and testing, all images were center-cropped and resized to a size of 256 × 256. Data augmentations, including random horizontal flips, vertical flips, random crops, and random rotations, were performed during training to prevent over-fitting. CycleGAN~\cite{cyclegan} and the method proposed in~\cite{9763342} were also trained with their official implementations for comparison. To maintain the fairness of all our comparisons, all models were trained with an RMSprop optimizer for 200 epochs with an initial learning rate of 
0.0001 for the discriminator and 0.00005 for the generator. The learning rate decayed by 10 every 100 epochs. The optimal $\lambda$ which yielded results for this paper was 40.  Our software package is available at \href{https://github.com/Retinal-Research/OTE-GAN}{https://github.com/Retinal-Research/OTE-GAN}.


\subsection{Evaluation Metrics}
\begin{figure}[!t]
\begin{minipage}[b]{1.0\linewidth}
  \centering
  \centerline{\includegraphics[width=8.5cm]{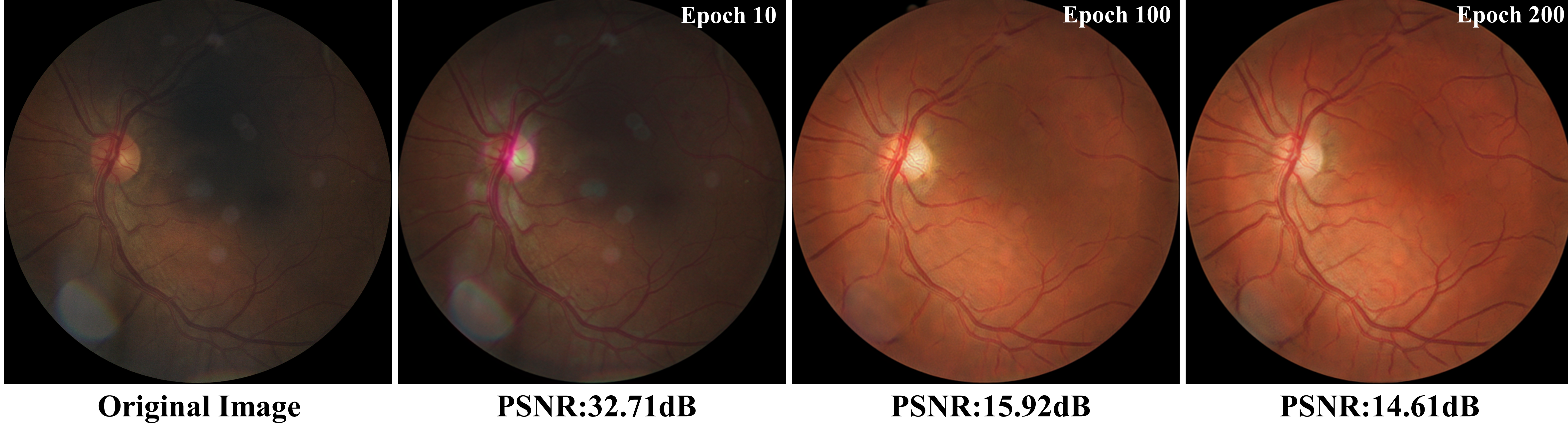}}
\end{minipage}
\caption{Visualization of the enhanced images during training at different epochs. High PSNR values do not guarantee improved denoising results.}
\label{fig:epoch_res}
\end{figure}

\noindent \textbf{No-Reference Quality Assessment.} 
Evaluating the quality of the enhancements without having access to the ground-truth high-quality images is challenging. Cheng et al.~\cite{9434005} measured the peak signal-to-noise ratio (PSNR) and structural similarity index measurement (SSIM) between the enhancements and the input low-quality images. While in our experiments, we found out that the PSNR between the low-quality images and their enhancements do not necessarily yield a high PSNR and SSIM, as shown in Fig.~\ref{fig:epoch_res}. We conducted two evaluation metrics to quantitatively assess the performance of the enhancements without knowing their ground-truth high-quality counterparts. 

The first no-reference quality evaluation metric is the Converted Ratio (CR) which is defined as the percentage of high-quality images among the enhancements. A ResNet-50 \cite{He_2016_CVPR} with efficient channel attention \cite{DBLP:journals/corr/abs-1910-03151} was trained on the EyeQ dataset with three labels (high quality, usable, and reject) to predict the quality of retinal images. Our trained CR evaluation model achieved a Cohen's kappa coefficient of 0.918 and an AU-ROC of 0.976 on the EyeQ testing set.

Secondly, a task-specific evaluation is performed on Diabetic Retinopathy prediction indicated by Classification Accuracy, Cohen's kappa coefficient (kappa), and Area under the receiver operating characteristic curve (AU-ROC). We performed the training on the enhanced images produced by three models, and also used them for model evaluation based on the EyeQ testing set. As shown in Table~\ref{tab:1}, our method outperformed the other two competitors in all evaluation metrics. All of our training setup for ResNet-50 using to evaluation follows the paper proposed in \cite{https://doi.org/10.48550/arxiv.2210.05946}, like data augmentation, learning rate, and scheduler.

\begin{figure}[!t]
\begin{minipage}[b]{1.0\linewidth}
  \centering
  \centerline{\includegraphics[width=\textwidth]{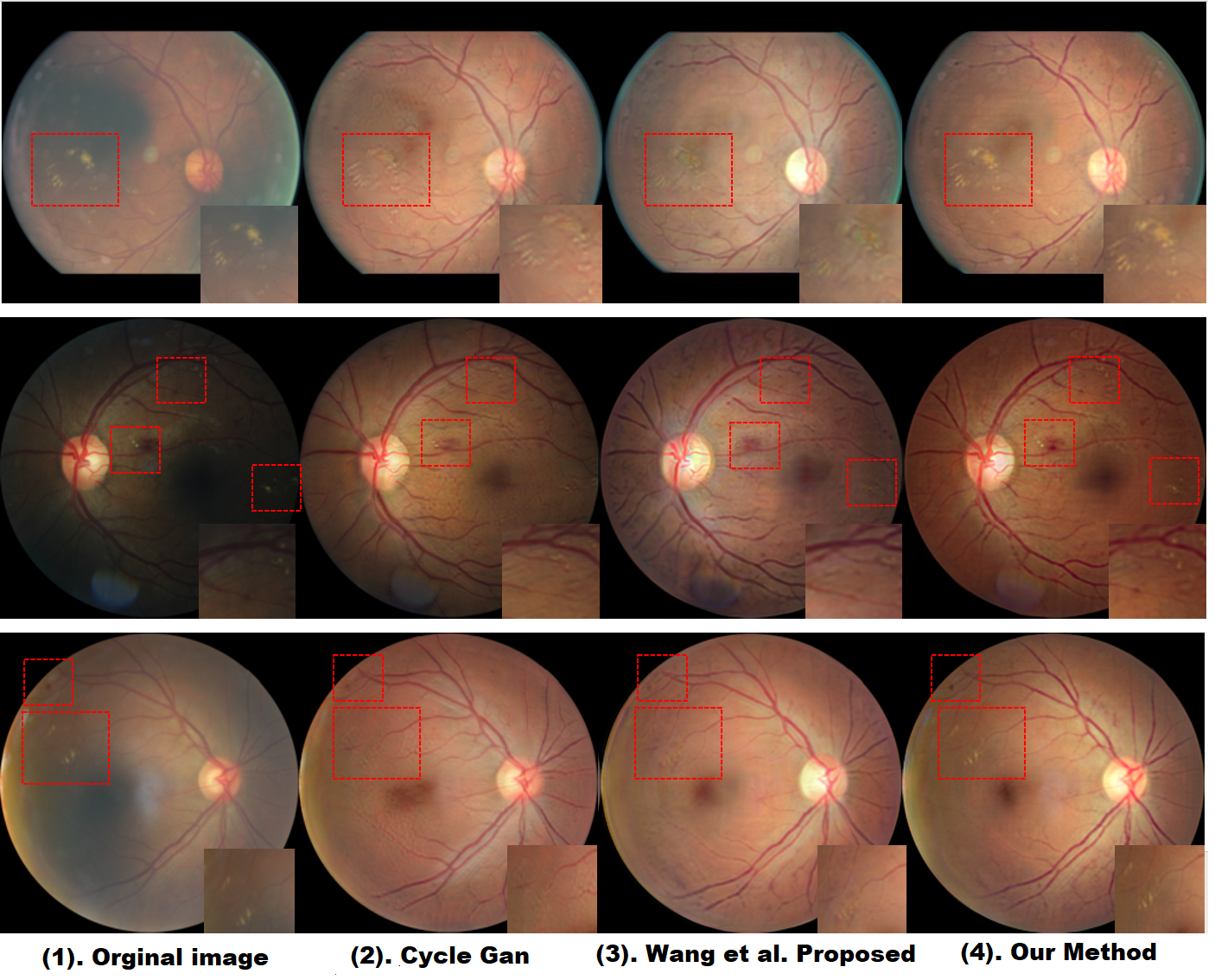}}
\end{minipage}
\vspace{-0.5cm}
\caption{Perceptual comparison of our proposed method to CycleGAN and OTT-GAN. Our proposed method provides better preservation in terms of lesion information and keeps vessel structure consistent with the original image.}
\label{fig:non_ref}
\end{figure}

\vspace{0.03cm}
Perceptually, CycleGAN and the OTT-GAN change the structure of vessels and smooth lesions to some extent, as shown in Fig.~\ref{fig:non_ref}. While our proposed method maintained the consistency of local structures of optic discs, vessels, and particularly lesions between low-quality images and their enhancements compared to the other two competitors.
\begin{table}[htp]
\vspace{-0.6cm}
\centering
\caption{Comparison No-Reference Evaluation Metrics.}\label{tab:1}
\begin{NiceTabular}{*{1}{p{2.1cm}}*{4}{c} }[hvlines]
\centering Method & CR & Accuracy & Kappa & AUC \\ 
\Block{1-1}{CycleGAN} & 0.2199 & 0.7148 & 0.5378  &  0.9083   \\
\Block{1-1}{Wang et al.} & 0.1835 & 0.6996 & 0.5105  & 0.8995    \\
\Block{1-1}{\textbf{Ours}} &  \textbf{0.2404} &  \textbf{0.7450} & \textbf{0.6349}  & \textbf{0.9255} \\
\end{NiceTabular}
\end{table}
\newline
\noindent \textbf{Full-Reference Quality Assessment.} 
We degraded 1400 high-quality images randomly selected from the EyeQ test dataset with the method proposed by Shen et al.~\cite{shen2020modeling} to simulate light interference, image blurring, and image artifacts.
We evaluated the performance of enhancement with a PSNR and SSIM between the degraded low-quality images and their high-quality counterparts. Our proposed method demonstrated superiority over the other two competitors statistically in terms of PSNR and SSIM, as shown in Table~\ref{tab:2}.

\vspace{-0.5cm}
\begin{table}[htp] 
\centering
\caption{Comparison Full-Reference Evaluation Metrics.}\label{tab:2}
\begin{NiceTabular}{*{1}{p{3cm}}*{2}{c} }[hvlines]
\centering Method & PSNR & SSIM \\ 
\Block{1-1}{Degration Images} & 17.2966 & 0.7399  \\
\Block{1-1}{CycleGAN} & 18.6820 & 0.7370   \\
\Block{1-1}{Wang et al.} & 18.7705 & 0.7335 \\
\Block{1-1}{\textbf{Ours}} &  \textbf{19.6757}  &  \textbf{0.7631}\\
\end{NiceTabular}
\end{table}
\vspace{-0.5cm}

\section{Conclusion}
In this work, we proposed an efficient and effective GAN-based unpaired image-to-image translation framework to tackle the blind enhancement of poor-quality retinal fundus images built on the optimal transport theory. To further enforce the consistency of local semantics (optical discs, vessel structures, lesions, etc) before and after the enhancement, we introduced a maximal mutual information consistency mechanism with a consistency loss, a lesion resampling strategy, in conjunction with a specific network design. Perceptual and quantitative no-reference and full-reference assessments on the EyeQ dataset showed the superiority of our proposed method over two main state-of-the-art competitors. Aside from the commonly used PSNR and SSIM image-quality evaluation metrics, we demonstrated the effectiveness of our proposed method on task-specific evaluation of diabetic retinopathy, suggesting the potential of our work in assisting real-world clinical diagnoses.


\section{Compliance with ethical standards}
\label{sec:ethics}
This research study was conducted retrospectively using human subject data available in open access by  \cite{10.1007/978-3-030-32239-7_6}. Ethical approval was not required as confirmed by the license attached with the open-access data.

\section{Acknowledgments}
\label{sec:acknowledgments}
This work was partially supported by the National Institutes of Health (R01EY032125, R21AG065942, R01EB025032, and R01DE030286), and the state of Arizona via the Arizona Alzheimer Consortium.

\bibliographystyle{IEEEbib}
\bibliography{refs}

\end{document}